\documentclass[twocolumn,aps,pre,reprint,superscriptaddress,longbibliography]{revtex4}
\usepackage{amsmath,amssymb}\usepackage{graphicx}\usepackage{subfigure}\usepackage{color} \usepackage{times}\usepackage{tabularx}

\begin{document}
\title{\bf Comment on ``Macroscopic surface charges from microscopic simulations'' [J. Chem. Phys. 153, 164709 (2020)]}
\author{Zhonghan Hu} \email{zhonghanhu@sdu.edu.cn}
\affiliation{Qingdao Institute for Theoretical and Computational Sciences (QiTCS), Shandong University, Qingdao, 266237, P. R. China}
\affiliation{Institute of Frontier and Interdisciplinary Science, Shandong University, Qingdao, 266237, P. R. China}
\maketitle 
Sayer and Cox (hereafter referred as the authors) recently presented a method to simulate a slab of electrolyte confined between walls composed of $n+1$ layers of charged surfaces, depicted in Fig.1 (b) of ref.\cite{Sayer_Cox2020}, under the 3-dimensional periodic boundary condition (PBC). The authors strongly suggested that
both $H_{\rm YB}$, which is the Yeh-Berkowitz (YB) method that adds a correction term to the usual Ewald sum with the tinfoil boundary condition (e3dtf)\cite{Yeh_Berkowitz1999}, and the mirrored slab geometry are unsuitable for modeling those systems. In our opinion, however, both YB method and the mirrored slab geometry can be
used to achieve the goal of ref.[1], as already explained in a previous publication\cite{Pan_Hu2019}. Besides, all methods that satisfy the symmetry-preserving mean-field (SPMF) condition\cite{Pan_Hu2019,Hu2014spmf} work well at an arbitrarily large value of $n$, not just $n=1$, as pointed out incorrectly by the authors. 
At last, while the slab system by itself is of interest to study, the authors' assertion of its physical relevance is completely wrong --- the corresponding Coulomb system in the absence of an external electric field does not really exist in the framework of statistical mechanics.

\begin{figure}[h!] \centerline{\includegraphics[width=3.0in]{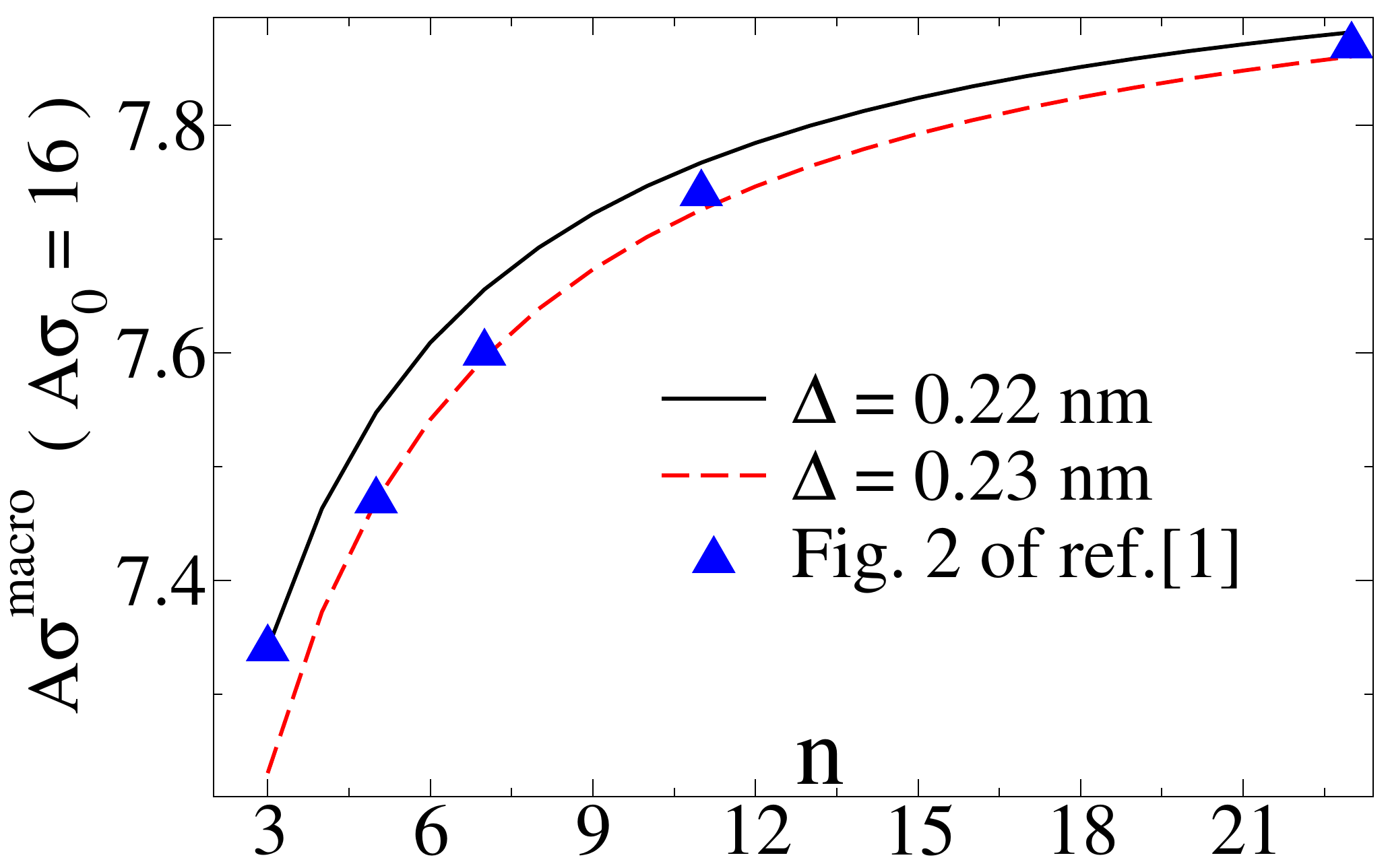}}
\caption{Spurious compensating charges as a function of $n$ in the presence of no external field given the smallest (black solid line) and largest (red dash line) possible values of $\Delta$. Data points (blue filled triangles) extracted from Fig.2 in ref.\cite{Sayer_Cox2020} are listed in Table~\ref{tab:data}.}
\label{fig:fig2}\end{figure}

I shall first simply repeat the theory in ref.\cite{Pan_Hu2019} for the present case of electrolyte (dielectric constant $\varepsilon_r = \infty$) to illustrate the artifacts produced by e3dtf in slab geometry. Physics at the level of high school states that a pair of infinitely large walls with the surface charge densities
$\pm \sigma$ separated at a finite distance of $d$ produces an electric field of $\sigma/\varepsilon_0$ inside the wall, and zero electric field outside the wall as a consequence of the $1/r$ Coulomb interaction. When such a pair of walls is simulated under the PBC of length $L$, the electric field produced by the corresponding
e3dtf potential, which is modified from $1/r$, is instead $ \sigma (L-d)/(\varepsilon_0 L) $ inside, and $-\sigma d/(\varepsilon_0 L) $ outside (see bottom panel of Fig.1 in ref.\cite{Pan_Hu2019}). 
In the slab system depicted in Fig.1 (b) of ref.\cite{Sayer_Cox2020}, there are $(n+1)/2$ pairs of fixed walls ($n+1$ solid lines), each of which is separated at a distance of $R$ with the surface charge densities $\mp\sigma_0$, and one pair of walls composed of mobile ions ($2$ dash lines) with the surface charge densities
$\pm\sigma^{\rm macro}$ separated at $ L - n R - \Delta$ after the left dash line at $x$ is PBC-transformed to the right at $L+x$ to obtain a continuous region of bulk electrolyte as done in Fig. 1 of ref.\cite{Pan_Hu2019}.  
The continuous bulk electrolyte therefore locates outside the $(n+1)/2$ pairs of fixed walls, and simultaneously inside the pair of mobile walls. At equilibrium, it experiences no net electrostatic forces on average:
\begin{equation}  - \frac{n+1}{2} \frac{\sigma_0}{\varepsilon_0}\frac{R}{L} + \frac{\sigma^{\rm macro}}{\varepsilon_0}\frac{nR + \Delta}{L}   = 0. \label{eq:e3dtfequi}\end{equation}
Relying on the fact that $\Delta$ depends weakly on the exerted electric field, which is explained in Fig. 2 of ref.\cite{Pan_Hu2019}, the derived compensating charge as a function of $n$,
\begin{equation} \sigma^{\rm macro} = \frac{\sigma_0}{2} \frac{n+1}{n+\Delta/R},  \label{eq:macro} \end{equation}  
therefore explains successfully the simulation data in ref.\cite{Sayer_Cox2020} as shown by Table~\ref{tab:data} and Fig.~\ref{fig:fig2}. 
A more elegant theory handling the general case of multiple dielectric fluids with arbitrary values of $\varepsilon_r$ ($1 \leqslant \varepsilon_r \leqslant \infty $) has been presented extensively in ref.\cite{Pan_Hu2019} to explain the otherwise not well explained simulation data of ref.\cite{Zhang_Sprik2016} (see Fig. 4 in ref.\cite{Pan_Hu2019} and Fig. 8 in ref.\cite{Zhang_Sprik2016}) where the finite-field method, appreciated deeply now by the authors, was originally presented.

\begin{table}[h!]\centering 
\begin{tabular}{cccccc} \hline
n                    \quad\quad &\quad    3     \quad &\quad     5     \quad &\quad     7     \quad &\quad     11     \quad &\quad     23    \\  \hline
$A\sigma^{\rm macro}$ \quad\quad &\quad  7.34    \quad &\quad    7.47   \quad &\quad   7.60    \quad &\quad    7.74    \quad &\quad    7.87   \\ \hline 
$\Delta ({\rm nm}) $ \quad\quad &\quad  0.22     \quad &\quad   0.23   \quad &\quad   0.23    \quad &\quad    0.23    \quad &\quad    0.23   \\ \hline  \end{tabular}
\caption{Simulation setup, $\sigma^{\rm macro}$ extracted from Fig. 2 of ref.\cite{Sayer_Cox2020}, and $\Delta$ derived from Eq.~\eqref{eq:macro} with $R=0.1628$ nm}\label{tab:data}  \end{table}

The existence of the nonzero compensating charges violating the basic argument of high school physics reflects essentially the spurious effect of the e3dtf potential for the slab geometry. Of course, there will be no compensating charges at all once the periodicity in the $z$ direction is removed and a formally exact Ewald2D
potential or its SPMF approximation is used subsequently (e.g. refs.\cite{Parry1975,Pan_Hu2014,Pan_Hu2019}).
Indeed, the mentioned YB method is an example of SPMF approximation to Ewald2D by adding a term corresponding to a planar limit of $k\to0$ term of the usual Ewald3D sum\cite{Hu2014ib,Pan_Hu2017},
\begin{equation} - \frac{1}{2\varepsilon_0V} \sum_{i<j} q_i q_j  z_{ij}^2 = \frac{1}{2\varepsilon_0 V} \left(\sum_j q_j z_j \right)^2 = \frac{V P^2}{2\varepsilon_0}, \label{eq:pib} \end{equation}
where $V$ is the volume of the primary unit cell and $P$ is the density of total dipole moment in the $z$ direction. $z_{ij}^2$ in the present setup depicted in Fig. 1(b) of ref.\cite{Sayer_Cox2020} is instead interpreted as the non-periodic counterpart of the minimum image convention (see Fig.5 of  ref.\cite{Pan_Hu2019}).

The above discussion for the case of no external field immediately suggest that, in the presence of an applied external electric field, there are two routes to reach any amount of desired compensating charge $\sigma_{\rm desired}$ (e.g. $\sigma_0/2$ or $\sigma_0/4$) in simulation: either by simply applying any SPMF-condition
satisfied method (e.g. refs.\cite{Parry1975,Yeh_Berkowitz1999,Zhang_Sprik2016,Hu2014spmf,Pan_Hu2019}) or applying a SPMF-condition not satisfied method (like the usual e3dtf) but paying enough attention to the actual response and the spurious effect as in Eq.~\eqref{eq:e3dtfequi}. 
For the former, the electric field applied simply reads \begin{equation} E_{\rm external}^{\rm spmf} + \frac{\sigma_{\rm desired}}{\varepsilon_0} = 0. \label{eq:extspmf} \end{equation}
A special case of the above equation would be using the YB correction method --- Eq.~\eqref{eq:pib} plus the usual e3dtf  under the electric field of $E_{\rm external}^{\rm spmf} $ --- which is fully equivalent to setting $E_{\rm external}^{\rm spmf}  = D$ in eq. (4) of ref.~\cite{Sayer_Cox2020} up to a constant of $D^2V
/(8\pi)$ and a prefactor of $1/(4\pi \varepsilon_0)$. $D$ used by the authors is actually an external field ! Any method in this category would be either the exact Ewald2D or its SPMF approximation. 
For the latter, paying enough attention to the spurious effect as in Eq.~\eqref{eq:e3dtfequi} yields
\begin{equation} E_{\rm external}^{\rm e3dtf} - \frac{n+1}{2} \frac{\sigma_0}{\varepsilon_0}\frac{R}{L} + \frac{\sigma_{\rm desired}}{\varepsilon_0}\frac{nR + \Delta}{L} = 0,  \label{eq:exte3dtf} \end{equation}
which however involves the static limit of the SPMF approximation\cite{Pan_Hu2019,Yi_Hu2017mf}. 
In a word, as discussed already in ref.\cite{Pan_Hu2019}, there are plenty of SPMF approximations or the static limits, that account for the desired responses of mobile charges,  including the one appreciated, the one pointed out incorrectly, those denied, and many others not mentioned by the authors.

For the mirror slab geometry depicted in Fig. 1(c) of ref.\cite{Sayer_Cox2020} with or without an external field, two routes exist again. For the former with YB method, because of the mirror slab and the large empty space, the minimum image convention is the same as its non-periodic counterpart, much less attention should be paid in the employment of $z_{ij}$\cite{Pan_Hu2019}.
For the latter case of the static limit of SPMF, the second term of Eq.~\eqref{eq:exte3dtf} will cancel the contribution from its mirror ( $\sigma_0$ versus $-\sigma_0$). Of course, in the presence of an external electric field, the exposed surface and its mirror face two oppositely charged ``compensating'' layers respectively,
one is with cations, the other with anions. Either or both would be of particular interest.

In contrast to what $D = E_{\rm external}^{\rm spmf}$ and the equivalence between the developed method in ref.\cite{Sayer_Cox2020} and the SPMF approximation clearly state --- the slab system in fact responds to an applied external electric field, the authors, however,  mysteriously regarded their Hamiltonian as a realization of a macroscopic crystal placed inside the electrolyte in the
absence of any applied external electric field (see bottom of Fig.1(a) in ref.\cite{Sayer_Cox2020}). Such an assertion of the physical relevance is simply inconsistent with the framework of statistical mechanics. 
To be more precise, let ${\mathbf R}_1, {\mathbf R}_2, \cdots {\mathbf R}_N $ and ${\mathbf r}_1,{\mathbf r}_2, \cdots, {\mathbf r}_M$ specify the positions of immobile atoms in the crystal and the degrees of freedom of mobile particles (ions and atoms in water molecules) respectively, with both $M$ and $N$ on the order of
$10^{23}$, with the use of Coulomb potential $1/r$ and any short-range bonded and non-bonded interactions in the absence of an external electric field, one just can not write down a Hamiltonian with its corresponding partition function producing compensating surface charge densities (e.g.  $\pm\sigma_0/2$ ) --- nothing
to do with interpreting the experiment, truly or falsely --- there is just no such a Hamiltonian as a function of the  $\sim 10^{23}$ specific variables in the framework of statistic mechanics.

I would like to thank Xuhui Huang, Chu Li, and Zhuo Liu at HKUST for stimulating discussions and the support from NSFC (grant no.21873037).

\end{document}